\documentclass[10pt,conference]{IEEEtran}

\IEEEoverridecommandlockouts
\usepackage{cite}
\usepackage{amsmath,amssymb,amsfonts}
\usepackage{algorithmic}
\usepackage{graphicx}
\usepackage{textcomp}
\usepackage{xcolor}
\usepackage[strict]{changepage}
\usepackage{framed}
\usepackage{wrapfig}
\usepackage{listings}
\usepackage{multirow}
\usepackage{framed}
\usepackage{multicol}
\usepackage{booktabs}
\usepackage{url}
\usepackage{makecell}
\usepackage{balance}

\newcommand{\mcode}[1]{$\tt #1$} 

\definecolor{lightgray}{rgb}{0.83, 0.83, 0.83}
\definecolor{formalshade}{rgb}{0.95,0.95,1}
\definecolor{darkblue}{rgb}{0.0, 0.0, 0.55}

\newenvironment{formal}{%
  \MakeFramed{\advance\hsize-\width\FrameRestore}%
  \noindent\hspace{-4.55pt}
  \begin{adjustwidth}{}{7pt}%
  \vspace{2pt}\vspace{2pt}%
}
{%
  \vspace{2pt}\end{adjustwidth}\endMakeFramed%
}

\def\BibTeX{{\rm B\kern-.05em{\sc i\kern-.025em b}\kern-.08em
    T\kern-.1667em\lower.7ex\hbox{E}\kern-.125emX}}
\begin{document}

\title{Migrating to GraphQL: A Practical Assessment}

\author{\IEEEauthorblockN{Gleison Brito, Thais Mombach, Marco Tulio Valente}
\IEEEauthorblockA{ASERG Group, Department of Computer Science, Federal University of Minas Gerais, Brazil\\
\{gleison.brito, thaismombach, mtov\}@dcc.ufmg.br}
}

\maketitle

\begin{abstract}
GraphQL is a novel query language proposed by Facebook to implement Web-based APIs. In this paper, we present a practical study on migrating API clients to this new technology. First, we conduct a grey literature review to gain an in-depth understanding on the benefits and key characteristics normally associated to GraphQL by practitioners. After that, we assess such benefits in practice, by migrating seven systems to use GraphQL, instead of standard REST-based APIs. As our key result, we show that GraphQL can reduce the size of the JSON documents returned by REST APIs in 94\% (in number of fields) and in 99\% (in number of bytes), both median results.
\end{abstract}


\section{Introduction}
\label{sec:intro}

GraphQL is a novel query language for implementing Web-based APIs~\cite{graphql2015}. Proposed by Facebook in 2016, the language represents an alternative to popular REST-based APIs~\cite{fieldingT02,fieldingT00,fielding:2000}, shifting from servers to clients the decision on the precise data returned by API calls.
 To illustrate the usage of the language, suppose the REST API currently implemented by arXiv, the popular preprint service maintained by Cornell University. This API includes a {\em search} endpoint that allows clients to retrieve metadata about preprints with a given title. The result is a complex and large JSON document, with at least 33 fields. However, clients might need only a few ones (e.g., only the paper's URL). Despite that, the mentioned endpoint returns all fields in a JSON document, which should be parsed by the clients. After that, the unneeded fields are discarded, although they have consumed server, client, and network resources. By contrast, suppose arXiv decides to support GraphQL. Using the language, clients formulate a simple call like this one:

\begin{lstlisting}
search(title: "A Solution of the P versus NP Problem"){
    pdfUrl
}
\end{lstlisting}

By means of this query, the client request a single field (\mcode{ pdfUrl}) of a preprint entitled ``{\em A Solution of the P versus NP Problem}''. The result is a JSON file with just this URL. Therefore, instead of receiving a full document, with 33 fields, the client receives exactly the single field it needs to process.

GraphQL is gaining momentum and it is now supported by important web services,  as the ones provided by GitHub and Pinterest~\cite{hartig2018}. Despite that, we have few studies investigating the real benefits of using GraphQL for implementing Web-based APIs. Therefore, in this paper we ask the following research questions: {\em (RQ1)} What are the  key characteristics and  benefits of  GraphQL? {\em (RQ2)} What are the main disadvantages of GraphQL? {\em (RQ3)}  When using GraphQL, what is the { reduction in the number of API calls} performed by clients? {\em (RQ4)} When using GraphQL, what is the {reduction in the number of fields} of the documents returned by  servers? {\em (RQ5)} When using GraphQL, what is the reduction in the size of the documents returned 
by  servers?
To answer {\em RQ1} and {\em RQ2}, we conduct a grey literature review, covering 28 popular Web articles (mostly blog posts) about GraphQL. Since the query language has just two years, we focus on  grey literature, instead of analysing  scientific papers, as usually recommended for emerging technologies~\cite{ogawa1991towards, garousi2016, garousi2017}. As a result, we confirmed two key characteristics of GraphQL: (a)  support to an {\em hierarchical data model}, which can contribute to reduce the number of endpoints accessed by clients; (b) support to {\em client-specific queries}, i.e.,~queries where clients only ask for the precise data they need to perform a given task. Motivated by these findings, we also assess the benefits achieved by GraphQL in terms of a reduction in the number of API calls ({\em RQ3}) and in the number of fields returned by service providers ({\em RQ4}). To answer these questions, we manually migrated five clients of the GitHub REST API to use the new GraphQL API provided by GitHub. We also implemented a GraphQL wrapper for two endpoints of arXiv's REST API and migrated two open source clients to use this wrapper. Finally, to answer {\em RQ5}, we reimplemented in GraphQL 14 queries used in seven recent empirical software engineering papers, published at two major software engineering conferences (ICSE and MSR).

Our contributions are twofold: (1) we reveal that GraphQL does not lead to a reduction in the number of queries performed by API clients in order to perform a given task, when compared to the number of required REST endpoints. For example, in our migration study, we migrated 29 API calls that access REST endpoints (distributed over seven clients) to 24 GraphQL queries, which therefore does not represent a major reduction; (2) we reveal that client-specific queries can lead to a drastic reduction in the size of JSON responses returned by API providers. On the median, in our study, JSON responses have 93.5 fields, against only 5.5 fields after migration to GraphQL, which represents a reduction of 94\%. In terms of bytes, we also measure an impressive reduction: from 9.8 MB (REST) to 86 KB (GraphQL). Altogether, our findings suggest that API providers should seriously consider the adoption of GraphQL. We also see space for tool builders and researchers, with interest on providing support and improving the state-of-the-practice on GraphQL-based API development.

The rest of this paper has seven sections. Section~\ref{sec:in_a_nutshell} provides a brief introduction to GraphQL. Section~\ref{sec:grey_literature} describes a grey literature review, covering popular Web articles about GraphQL. Section~\ref{sec:migration} presents the migration study conducted to answer {\em RQ3} and {\em RQ4}. Section~\ref{sec:runtime} describes a study to evaluate the runtime gains achieved by GraphQL (and therefore answer {\em RQ5}).
Threats to validity are discussed in Section~\ref{sec:threats}; and related work is discussed in Section~\ref{sec:related-work}. Section~\ref{sec:conclusion} concludes the paper and summarizes lessons learned.

\section{GraphQL in a Nutshell}
\label{sec:in_a_nutshell}

This section introduces the key concepts of GraphQL. The goal is to make this paper self-contained; for a detailed and throughout presentation of the language we refer the reader to its documentation~\cite{graphql2015}.
Essentially, GraphQL allows clients to query a database represented by a {\em schema}. Therefore, this design represents a major shift from REST APIs: in REST, server applications implement a  list of endpoints (operations) that can be called by clients; by contrast, in GraphQL servers export a database, which can be queried by clients.
A GraphQL database is defined by a {\em schema}, which is a multi-graph~\cite{hartig2017}. In this multi-graph,  nodes are {\em objects}, which define a type and include a list of {\em fields}; each field has a name and also a type. Edges appear when objects define fields whose types are other object types.\footnote{Since an object $T_1$ can have multiple fields of type $T_2$, multiple edges can connect $T_1$ to $T_2$, leading to a multi-graph.} GraphQL proposes a simple DSL for defining schemas. To illustrate the usage of this language, we 
use a simple blogging system, with two object types: \mcode{Post} and \mcode{Author}. As presented in Listing~\ref{lst:schema},  object types are defined using the keyword \mcode{type}. In this example, \mcode{Post} has four fields: {\em id}, {\em author}, {\em title}, and {\em body}. The first field is a non-null \mcode{String} (the {\sc !} after the type discards \mcode{null} values). The second field ({\em author}) references another object type in the schema, called \mcode{Author} (lines 8-12). The remaining two fields in \mcode{Post} have a \mcode{String} type. Finally, schemas usually include a predefined type, called \mcode{Query}, which represents the entry point of GraphQL APIs (lines 14-16). A \mcode{Query} object exposes the object types that can be queried by clients and the arguments that must be provided by them. For example, \mcode{post} (line 15) is a query that accepts a non-null string as argument and returns the \mcode{Post} object having this string as \mcode{id}.

\begin{lstlisting}[caption=Schema with two  types ({\em Post} and {\em Author}) and a {\em Query} end-point, label=lst:schema]
type Post {
  id: String!
  author: Author
  title: String
  body: String
}

type Author {
  id: String!
  name: String
  email: String
}

type Query {
  post(id: String!): Post
}  
\end{lstlisting}

GraphQL also defines a query language, used by clients. Listing~\ref{lst:query-examples} shows three examples of queries in this language. In the first query (\mcode{PostByTitle}), the client asks for the \mcode{Post} object with \mcode{id} equals to 1000; specifically, the client only requests the \mcode{title} field of this object.  The second query (\mcode{PostByTitleAndBody}) is similar, but in this case the client asks for two fields, \mcode{title} and \mcode{body}. Finally, the last query (\mcode{PostByTitleAndAuthor}) asks for the \mcode{title} and \mcode{author} of the same \mcode{Post}. Since \mcode{author} is another object, we have to specify its queried fields, in this case only \mcode{name}. The result of this third query is presented in Listing~\ref{lst-PostByTitleAndUser-Result}. As we can see, the result is a JSON object, which should be parsed and possibly deserialized by clients.

\begin{lstlisting}[caption=Querying distinct data from a given {\em Post} object, label=lst:query-examples]
query PostByTitle{
    post(id:"1000"){
      title
    }
}
query PostByTitleAndBody{
    post(id:"1000"){
      title
      body
    }
}
query PostByTitleAndAuthor{
    post(id:"1000"){
      title
      author {
        name
      }
    }
}
\end{lstlisting}

\begin{lstlisting}[caption=JSON object returned by \mcode{PostByTitleAndAuthor} query, label=lst-PostByTitleAndUser-Result]
{ "data": {
    "post":{
       "title": "GraphQL: A data  query language"
       "author":{
          "name": "Lee Byron"
       }
    }
  }
}
\end{lstlisting}

To respond to queries, the developer of a GraphQL server must implement a {\em resolver} function for each query declared in the \mcode{Query} type. These functions are called each time the GraphQL server engine needs to retrieve an object type specified in a query. Typically, these functions retrieve these objects from an underlying data structure, which can be any kind of database (relational, non-relational, in-memory, etc).
Finally, it is also possible to define another predefined type in schemas, called \mcode{Mutation}, which is used to insert new objects on the server's database or modify existing ones. Listing~\ref{lst-mutation-example} shows an example that defines an \mcode{addPost} mutation, which receives a \mcode{Post} object (and returns the object, just to confirm the operation has been successfully executed). Each endpoint (operation) in a \mcode{Mutation} type must have a resolver function, which implements the operation. In our running example, this function must insert the \mcode{Post} object received as argument in the underlying database.

\begin{lstlisting}[caption=Mutation operation for {\em Post} objects, label=lst-mutation-example]
type Mutation {
   addPost(post: Post): Post
}
\end{lstlisting}

\section{Grey Literature Review}
\label{sec:grey_literature}

This section reports the results of a systematic analysis of the grey literature about GraphQL, covering documents and discussions on blogs, tutorials, and similar Web articles. Our goal is to better understand the key characteristics, benefits, and shortcomings appointed by practitioners who had a real experience with the language. Since it is a new technology, papers about GraphQL are not common in the scientific literature. Therefore, for such emerging technologies, a grey literature tends to provide a better coverage of relevant documents than a traditional literature review~\cite{garousi2016,barik2015}.\footnote{Nevertheless, peer-reviewed articles are also discussed in this paper, but in Section~\ref{sec:related-work} (Related Work).}

\subsection{Study Design}

\begin{figure}[h]
\centering
\includegraphics[width=0.49\textwidth]{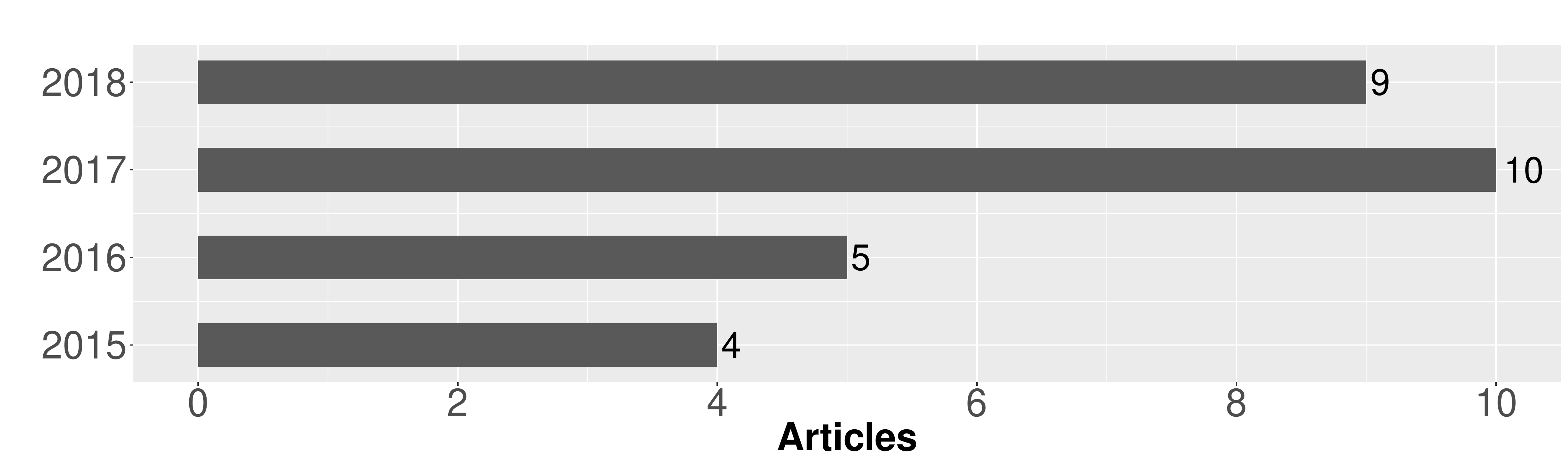}
\caption{Grey literature articles by year of appearance on Hacker News}
\label{fig:articlesOverYears}
\end{figure}

\begin{figure}[h]
\centering
\includegraphics[width=0.24\textwidth]{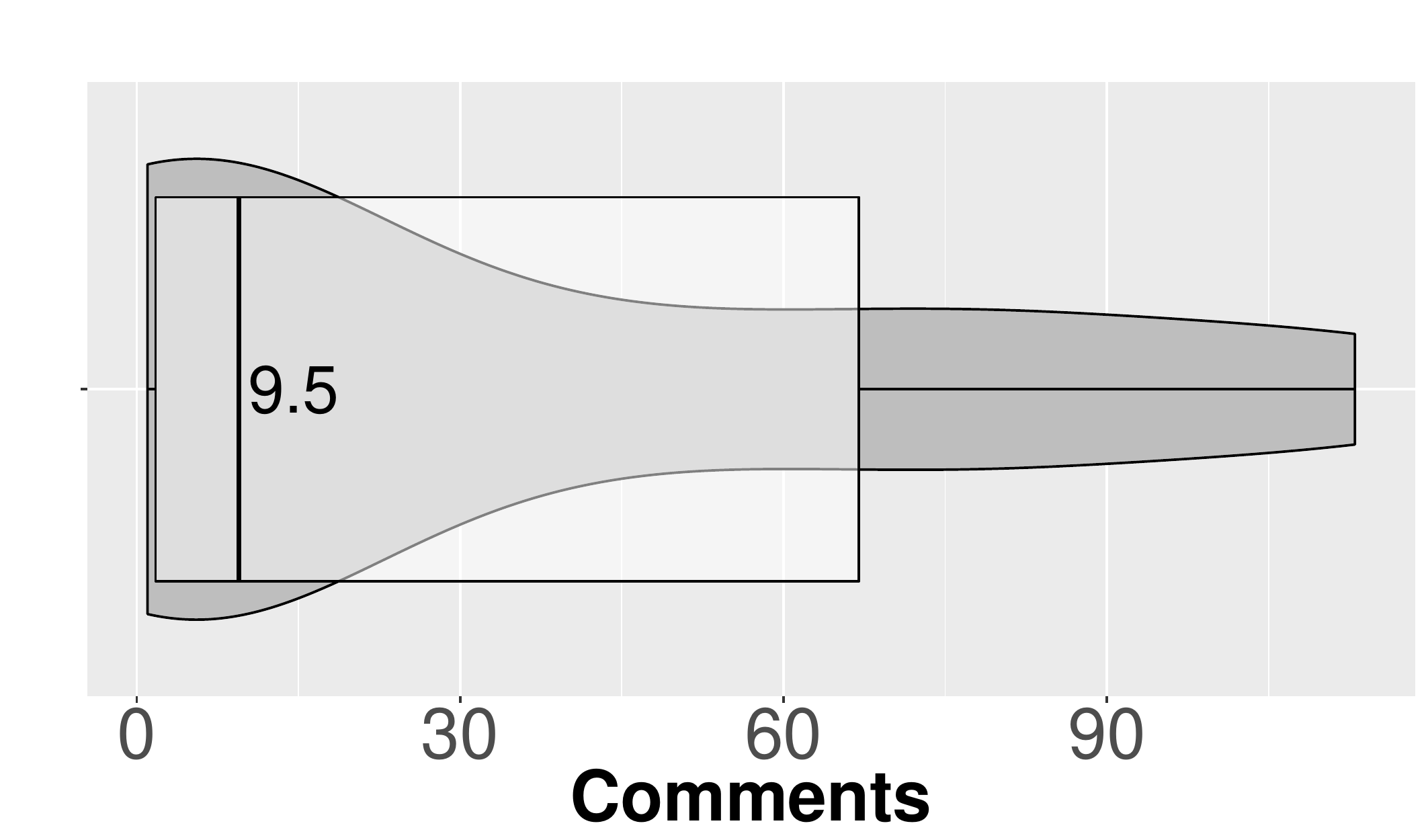}
\includegraphics[width=0.24\textwidth]{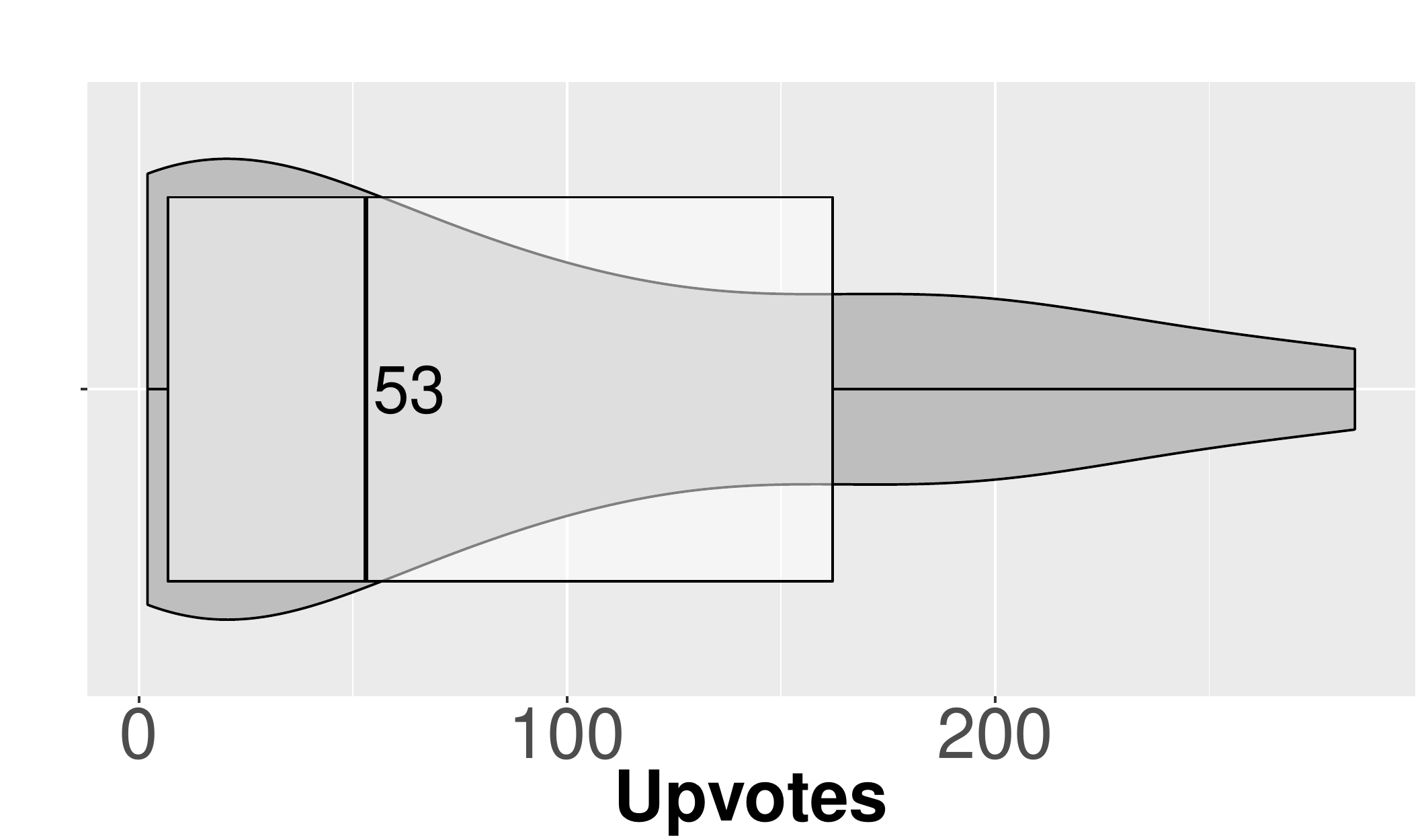}
\caption{Number of comments and upvotes on Hacker News (for the articles included in the grey literature review)}
\label{fig:articleMetrics}
\end{figure}

To retrieve an initial list of Web articles considered in this review, we used Hacker News,\footnote{\url{https://news.ycombinator.com}} which is a news aggregator site widely used by practitioners~\cite{aniche2018}. Recently, other similar reviews have used Hacker News as data source, e.g.,~a grey literature review on clouding computing services~\cite{leitner2018}. To retrieve Hacker News documents, we used the Algolia search engine,\footnote{\url{https://hn.algolia.com}} querying for posts containing \textit{graphql} in their titles, as in September, 2018. We found 1,242 articles. We then sequentially removed articles that do not include a valid URL (286 articles), that do not contain comments (760 articles), or that are just promoting a tool or project (168 articles). After this filtering step, we selected 28 articles for analysis (1,242 - 286 - 760 - 168), which we refer  as \textit{A1} to \textit{A28}.\footnote{Detailed information at  \url{https://github.com/gleisonbt/migrating-to-graphql}} Figure~\ref{fig:articlesOverYears} shows the year when these articles appeared on Hacker News. We can see an increase in the interest on GraphQL in the last four years. Interestingly, we found four articles in 2015, i.e; before GraphQL official release.
Figure~\ref{fig:articleMetrics} shows violion plots with the distribution of the number of comments and upvotes of the 28 articles. The median number of comments is 9.5; and the median number of upvotes is 53 (which is usually enough to put the article in the front page of Hacker News).

After collecting the articles, the first author of this paper carefully read them and followed an open coding protocol to provide answers to the first two research questions:\\[-0.25cm]

\noindent {\em RQ1}: What are the  {\em characteristics} and {\em benefits} of  GraphQL?\\[-0.35cm]

\noindent {\em RQ2}: What are the main {\em disadvantages} of GraphQL?


\subsection{Results}

\noindent{\em RQ1: Key Characteristics and Benefits}\\[-0.2cm]

\noindent\emph{GraphQL is strongly typed}, since all objects and fields have types (as mentioned in A1, A2, A5, A6, A10, and A28). This  contributes to better tooling support, as reported in this article:
\begin{formal} \em \small
GraphQL is strongly-typed. Given a query, tooling can ensure that the query is syntactically correct and valid within the GraphQL type system before execution. (A5)
\end{formal}

A related benefit is the possibility of having  better error messages, e.g.,~{\em [types] allow GraphQL to provide descriptive error messages before executing a query} (A28).\\[-0.1cm]

\noindent{\em GraphQL enables client-specified queries}, as mentioned by almost half of the articles (A1, A2, A5, A6, A10, A14, A17, A19, A21, A25, A26, and A27). The following article nicely describes this characteristic:

\begin{formal} \em \small
In GraphQL, the specification for queries are encoded in the client rather than the server. These queries are specified at field-level granularity. In the vast majority of applications written without GraphQL, the server determines the data returned in its various scripted endpoints. A GraphQL query, on the other hand, returns exactly what a client asks for and no more. (A5)
\end{formal}

This characteristic makes GraphQL particularly interesting for mobile applications, which often face limited bandwidth and speed (A4, A5, A14, and A26). It also {\em moves the focus of development to client apps, where designers and developers spend their time and attention} (A1). Finally, client-specific queries allow servers to better understand the needs of clients (A9, A12) and therefore improve the quality of their service:

\begin{formal} \em \small
It's great for service operators too, because its explicitness allows them to get a better understanding of exactly what their users are trying to do. (A9)
\end{formal}

\noindent\emph{GraphQL data model is hierarchical}, as mentioned in five articles (A1, A2, A3, A5, and A8) and defined as:

\begin{formal} \small
\textit{As its name would suggest, GraphQL models objects as a graph. Technically, the graph starts with a root node that branches into query and mutation nodes, which then descend into API-specific resources. (A2).}
\end{formal}

 This characteristic allows clients to retrieve data from multiple sources (or endpoints) in a single request,  therefore acting as  gateways for different APIs (A3, A4, A14, and A20):
\begin{formal} \small
\emph{GraphQL makes it easy to combine multiple APIs into one, so you can implement different parts of your schema as independent services.} (A20)
\end{formal}

\noindent\emph{Introspection}, which allows clients to inspect the types and fields defined in a schema, at runtime (A1, A3, A9, A16, and A28). Combined with a static type system, introspection allows clients to {\em learn and explore an API quickly without grepping the codebase or wrangling with cURL.} (A28). It also frees servers to support an interface description language, which are usually not available for REST; and when available they are {\em often not completely accurate because the description is not tied directly to the implementation} (A28).\\[-0.1cm]

\noindent\emph{Deprecation}: As common in mainstream programming languages, it is possible to deprecate fields, using a \mcode{@deprecated} annotation (A1, A2 and A19). However, in GraphQL, new fields added to a type do not lead to breaking changes (as in standard APIs~\cite{saner2018,saner2017-laerte}); and deprecations can be restricted to deleted fields. As a result, the pressure for versioning is less frequent, as mentioned in this article:

\begin{formal} \em \small
This process removes the need for incrementing version numbers. We still support three years of released Facebook applications on the same version of our GraphQL API. (A1)
\end{formal}

\noindent{\em RQ2: Disadvantages}\\[-0.2cm]

\noindent\textit{GraphQL does not support information hiding}. GraphQL does not support private fields, i.e., all fields are visible to client applications (A8, A11, A18, A20 and A24). Furthermore, according to A18, GraphQL queries tend to be more complex to implement, since they require a detailed understanding of the data schema, which can be a time-consumming task in large APIs:

\begin{formal} \em \small
By design, a developer who integrates against GraphQL needs to know the names of the fields to access, their relation to other objects and when to retrieve them. (A18)
\end{formal}

\noindent{\it Complex caching}:  
In GraphQL, each query can be different, even though operating on the same type. This demands more sophisticated server-side caching, as mentioned in this article:

\begin{formal} \em \small
GraphQL does not follow the HTTP specification for caching and instead uses a single endpoint. Thus, it's up to the developer to ensure caching is implemented correctly \ldots (A20)
\end{formal}

\noindent\textit{Performance: } GraphQL servers can have to process complex queries (e.g.,~queries with deep nesting) that can consume server resources (A8, A11, A20, A23, and A25), as mentioned in the following article:

\begin{formal} \small
\textit{Great care has to be taken to ensure GraphQL queries don't result in expensive join queries that can bring down server performance or even DDoS the server. (A20)}
\end{formal}

Figure 3 summarizes the grey literature review results, by presenting the key characteristics, benefits, and disadvantages of GraphQL, and the number of articles mentioning them.


\begin{figure}[ht!]
\centering
\includegraphics[width=0.49\textwidth]{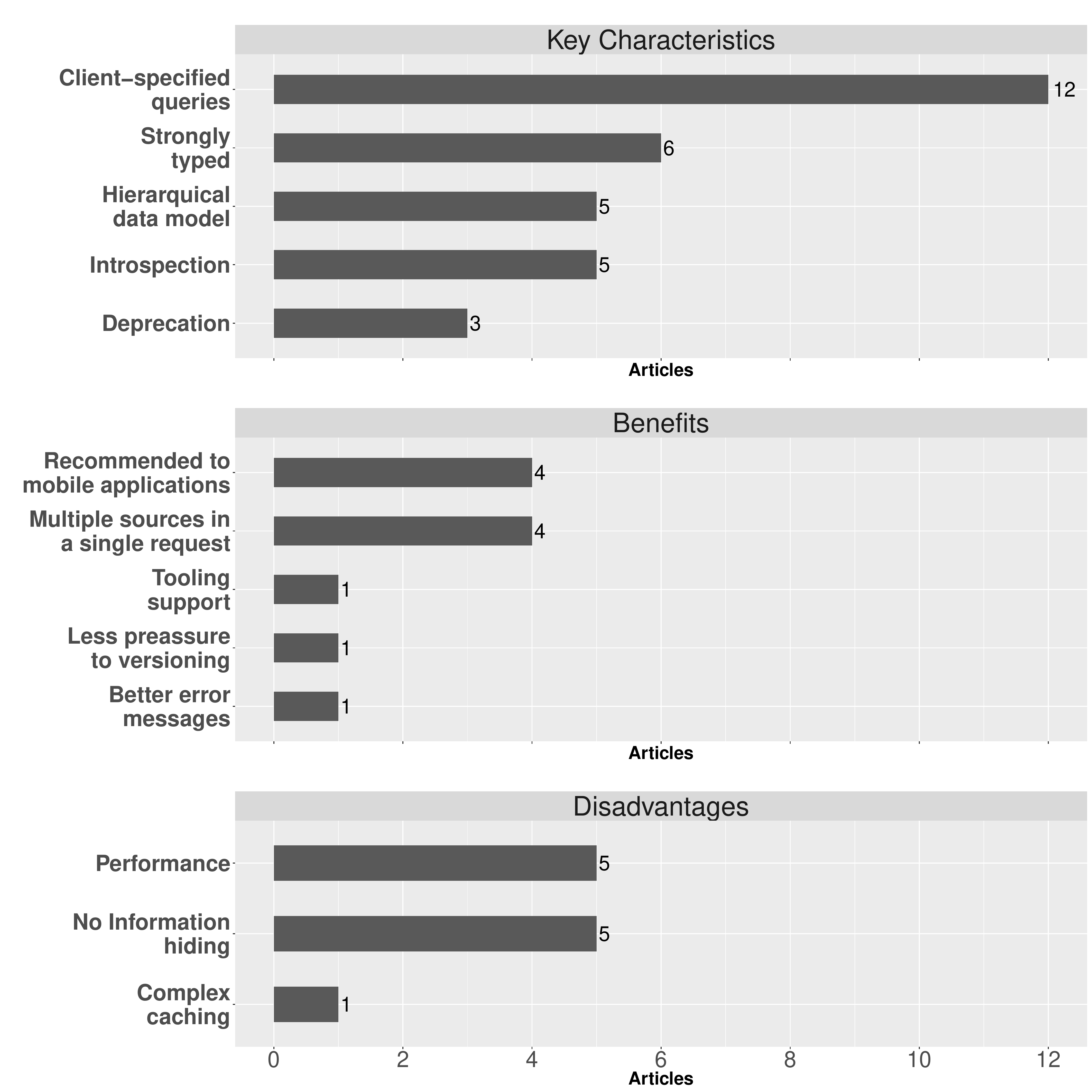}
\caption{Summary of grey literature findings}
\label{fig:summaryGreyLiterature}
\end{figure}

\section{Migration Study}
\label{sec:migration}

With this second study, we aim to quantitatively evaluate two key characteristics associated to GraphQL in the  grey literature: (1) clients can precisely request the data they need from servers (due to the support to client-specific queries) (2) clients rely on a single endpoint to retrieve the data they need (due to a hierarchical data model). 
In the study, we migrate seven client applications based on REST APIs to GraphQL. Then, we assess the gains achieved by the GraphQL version. Specifically, we answer two research questions:\\[-0.3cm]

\noindent {\em RQ3: When using GraphQL, what is the reduction in the number of API calls performed by clients?} GraphQL clients normally implement a single query to retrieve all data they need to perform a given task; by contrast, when using REST, clients frequently have to access multiple endpoints. Therefore, in this RQ, we compare the number of endpoints accessed by REST clients with the number of endpoints accessed by the same clients after refactored to use GraphQL.\\[-0.3cm]

\noindent {\em RQ4: When using GraphQL, what is the reduction in the number of fields of the JSON documents returned by  servers?} In GraphQL, client-specific queries  allow developers to inform precisely  the fields they need from servers. Therefore, in this RQ, we compare the number of fields in the following JSON documents: (a) returned by servers when responding to requests performed by  REST clients; (b) returned by servers when responding to queries performed by the same clients but after being migrated to use GraphQL.

\subsection{Study Design}

\noindent{\em Selected APIs:} First, to answer the proposed research questions, we selected the APIs provided by two widely popular services: GitHub and arXiv.\footnote{\url{https://arxiv.org}} GitHub is an interesting case study because the system provides both REST and GraphQL APIs (the latter since 2016). Moreover, GitHub's GraphQL API is quite complete and large, including 120 object types,
21 queries, and 62 mutations. By contrast, arXiv only provides a REST API. Therefore, we  implemented ourselves a small GraphQL API for the system, in the form of a wrapper for the original API. This wrapper supports two queries (\mcode{getPreprint} and \mcode{search}), as presented in Listing~\ref{lst-QueryTypeArxiv}. The first query (\mcode{getPreprint}) returns metadata about a preprint, given its ID. This metadata includes the paper's title, authors, DOI, summary, URL, etc (see Listing~\ref{lst-EntryTypeArxiv}).
The second query searches for preprints whose title match a given string; it is also possible to define the maximal number of results the query should return, the first result that should be returned and the sort order.
The implemented GraphQL wrapper was installed in a private server, in our research lab.\\

\begin{lstlisting}[caption=arXiv \mcode{Query}, label=lst-QueryTypeArxiv]
type Query {
   getPreprint(id: ID!): Preprint
   search(query: String!, maxResults: Int!,
          start: Int!, sortBy: String,
          sortOrder: String): [Preprint]
}
\end{lstlisting}

\begin{lstlisting}[caption=\mcode{Preprint} type, label=lst-EntryTypeArxiv]
type Preprint {
   id: ID
   pdfUrl: String
   published: String
   arxivComment: String
   title: String
   authors: [String]
   arxivUrl: String
   doi: String
   tags: [Tag]
   arxivPrimaryCategory: ArxivPrimaryCategory
   updated: String
   summary: String
}
\end{lstlisting}

\noindent\textit{Selected Clients:}
When searching for GitHub API clients, we first found that they usually have the tag (or topic) {\it GitHub}. Therefore, we selected five projects with this tag and that have at least 100 stars, as described in Table~\ref{tab:projDescr}. In the case of arXiv, we selected two clients mentioned in the project's page\footnote{\url{https://arxiv.org/help/api/index}} and that have their source code publicly available on GitHub (see also their names in Table~\ref{tab:projDescr}). Table~\ref{tab:selProjects} shows information about the programming language, number of stars, size (in lines of code), and contributors of the selected systems. The smallest project is \textsc{bibcure/arxivcheck} (131 LOC, one contributor, and five stars); the largest projects are \textsc{vdaubry/github-awards} (35,153 LOC, 15 contributors, and 1,296 stars) and \textsc{donnemartin/gitsome} (17,273 LOC, 24 contributors and 5,913 stars).\\[-0.2cm]

\begin{table}[ht]
\caption{Selected Projects}
\centering
\begin{tabular}{@{}l@{\hspace*{4pt}}p{5.4cm}@{}}
\toprule
\multicolumn{1}{c}{\textbf{Project}}                                       & \multicolumn{1}{c}{\textbf{Description}}                       \\ \midrule
\textsc{donnemartin/viz}                                                            & Visualization of GitHub repositories                  \\
\textsc{donnemartin/gitsome}                                                        & Command line interface for GitHub                       \\
\textsc{csurfer/gitsuggest}                                                         & A tool to suggest GitHub repositories \\
\textsc{guyzmo/git-repo}                                                            & Command line interface to manage Git services           \\
\textsc{vdaubry/github-awards}                                                      & Ranking of GitHub repositories                          \\
\textsc{bibcure/arxivcheck}                                                         & A tool to generate \BibTeX{} of arXiv preprints           \\
\begin{tabular}[c]{@{}l@{}}\textsc{karpathy}/\\\textsc{arxiv-sanity-preserver}\end{tabular} & Web interface for searching arXiv submissions            \\
\bottomrule
\end{tabular}
\label{tab:projDescr}
\end{table}

\begin{table}[ht]
\caption{Stats of Selected Projects}
\centering
\begin{tabular}{llrrr}
\toprule
\multicolumn{1}{c}{\textbf{Project}}                & \textbf{Lang.} & \multicolumn{1}{l}{\textbf{Stars}} & \multicolumn{1}{l}{\textbf{LOC}} & \multicolumn{1}{l}{\textbf{Contrib}} \\ \midrule
\textsc{donnemartin/viz}                 & Python            & 627                                 & 9,556                               & 1                                        \\
\textsc{donnemartin/gitsome}             & Python            & 5,913                                 & 17,273                               & 24                                        \\
\textsc{csurfer/gitsuggest}              & Python            & 613                                 & 389                               & 2                                        \\
\textsc{guyzmo/git-repo}                 & Python            & 764                                 & 5,602                               & 17                                        \\
\textsc{vdaubry/github-awards}           & Ruby              & 1,296                                 & 35,153                               & 15                                        \\ 
\textsc{bibcure/arxivcheck}              & Python            & 5                                 & 131                               & 1   \\ 
\begin{tabular}[c]{@{}l@{}}\textsc{karpathy/}\\\textsc{arxiv-sanity-preserver}\end{tabular} & Python            & 2,322                                & 2,431                               & 19  \\\bottomrule
\end{tabular}
\label{tab:selProjects}
\end{table}

\noindent\textit{Migration Step:} After selecting the APIs and client projects, the paper's first author exhaustively searched the code of each client looking for REST calls. He then migrated each one to use GraphQL. Just to show one example of migration,  in \textsc{csurfer/gitsuggest} the following REST endpoint is used to search GitHub for repositories matching a given string:

\begin{lstlisting}[]
GET /search/repositories
\end{lstlisting}

This endpoint requires three parameters: \mcode{q} (a string with the search keywords), 
\mcode{sort} (the sort field, e.g., {\em stars}), and the 
\mcode{order} ({\em asc} or {\em desc}).\footnote{\url{https://developer.github.com/v3/search/}} The request returns a JSON document with 94 fields, containing data about a repository. However, only three fields are used by \textsc{csurfer/gitsuggest}: owner's login, {\em description}, and {\em stargazers\_count}. Therefore, we changed the function that implements the search call to use the following GraphQL query, which retrieves exactly the three fields used by \textsc{csurfer/\-gitsuggest}:

\begin{lstlisting}[caption=Example of GraphQL query (\textsc{csurfer/gitsuggest}), label=lst-queryExample]
query searchRepos{
  search(query:$query, type:REPOSITORY, first: 100){
    nodes{
      ... on Repository{
                nameWithOwner
                description
                stargazers{
                   totalCount
                }
             }
      }
    }
}
\end{lstlisting}

\begin{table}[!h]
\caption{REST Calls}
\label{tab:restCalls}
\centering
\scriptsize
\begin{tabular}{@{}l@{\hspace*{1pt}}c@{\hspace*{2pt}}l@{\hspace*{1pt}}@{}}
\toprule
\multicolumn{1}{c}{\textbf{Project}}         & \multicolumn{1}{c}{\textbf{Func}} & \multicolumn{1}{c}{\textbf{REST endpoints}} \\ \midrule
\multirow{5}{*}{\textsc{csurfer/gitsuggest}}                                        & \multirow{4}{*}{F1}                   & \texttt{GET /users/:user}                        \\
                                                                           &                                       & \texttt{GET /users/:user/starred}                \\
                                                                           &                                       & \texttt{GET /users/:user/following}              \\
                                                                           &                                       & \texttt{GET /users/:user/starred}                \\ \cmidrule{2-3} 
                                                                           & F2                                    & \texttt{GET /search/repositories}                \\ \midrule
\multirow{11}{*}{\textsc{donnemartin/gitsome}}                                      & F3                                    & \texttt{GET /users/:user/followers}              \\ \cmidrule{2-3} 
                                                                           & F4                                    & \texttt{GET /users/:user/following}              \\ \cmidrule{2-3} 
                                                                           & F5                                    & \texttt{GET /repos/:owner/:repo/issues}          \\ \cmidrule{2-3} 
                                                                           & \multirow{2}{*}{F6}                   & \texttt{GET /users/:user/repos}                  \\
                                                                           &                                       & \texttt{GET /repos/:owner/:repo/pulls}           \\ \cmidrule{2-3} 
                                                                           & F7                                    & \texttt{GET /users/:user/repos}                  \\ \cmidrule{2-3} 
                                                                           & F8                                    & \texttt{GET /search/issues}                      \\ \cmidrule{2-3} 
                                                                           & F9                                    & \texttt{GET /search/repositories}                \\ \cmidrule{2-3} 
                                                                           & F10                                   & \texttt{GET /users/:user/starred}                \\ \cmidrule{2-3} 
                                                                           & \multirow{2}{*}{F11}                  & \texttt{GET /users/:user}                        \\
                                                                           &                                       & \texttt{GET /users/:user/repos}                  \\ \midrule
\multirow{6}{*}{\textsc{guyzmo/git-repo}}                                           & F12                                   & \texttt{GET /users/:user/repos}                  \\ \cmidrule{2-3} 
                                                                           & F13                                   & \texttt{GET /users/:user/gists}              \\ \cmidrule{2-3} 
                                                                           & \multirow{2}{*}{F14}                  & \texttt{GET /repos/:owner/:repo}                 \\
                                                                           &                                       & \texttt{GET /repos/:owner/:repo/pulls}           \\ \cmidrule{2-3} 
                                                                           & F15                                   & \texttt{GET /repos/:owner/:repo}                 \\ \cmidrule{2-3} 
                                                                           & F16                                   & \texttt{GET /repos/:owner/:repo}                 \\ \midrule
\multirow{2}{*}{\textsc{donnemartin/viz}}                                           & F17                                   & \texttt{GET /users/:user}                        \\ \cmidrule{2-3}  
                                                                           & F18                                   & \texttt{GET /search/repositories}                \\ \midrule
\multirow{3}{*}{\textsc{vdaubry/github-awards}}                                     & F19                                   & \texttt{GET /repos/:owner/:repo}                 \\ \cmidrule{2-3} 
                                                                           & \multirow{2}{*}{F20}                  & \texttt{GET /users/:user}                        \\
                                                                           &                                       & \texttt{GET /users/:user/repos}                  \\ \midrule
\textsc{bibcure/arxivcheck}                                                         & F21                                   & \texttt{GET /query/:search\_query}                                  \\ \midrule
\begin{tabular}[c]{@{}l@{}}\textsc{karpathy/}\\\textsc{arxiv-sanity-preserver}\end{tabular} & F22                                   & \texttt{GET /query/:search\_query}                                  \\ \bottomrule
\end{tabular}
\end{table}

In Listing~\ref{lst-queryExample}, the \mcode{search} query returns an union type, which might be either a \mcode{Repository}, \mcode{User}, or an \mcode{Issue} type, depending on the \mcode{type} argument. We use a feature of GraphQL called \textit{inline fragments} to access only the fields of the  \mcode{Repository} variant type. This variant is labeled as \mcode{... on Repository} (line 4).
Therefore, in this case one REST endpoint is replaced by one GraphQL query (RQ3's answer) and 91 fields (= $94 -3$) are retrieved but not used by the REST code (RQ4's answer). 

In total, the first author migrated 29 REST endpoint calls---distributed over the seven projects (see Table~\ref{tab:restCalls})---to use GraphQL queries. For the sake of legibility, we use labels F1 to F22 to refer to the functions including these REST calls (instead of the functions' original names). This migration effort consumed around 60 working hours (of the paper's first author), including the time to understand the clients code.\\[-0.2cm]

\noindent{\em Number of JSON fields:} To answer RQ4, we have to compute the number of fields returned by the original API calls (performed using REST) and by the migrated calls (using GraphQL). First, it is important to highlight that we only count root nodes, i.e., the ones that have a primitive value associated to them, instead of referring to another JSON entry. Second, when the returned fields are lists, we only consider a single list element. For example, Listing~\ref{lst-getUserFollowing} shows a JSON object that contains a list of users followed by a given GitHub user. The list contains three \mcode{nodes} elements, delimited by square brackets (lines 5-7). Each node contains only one root field called \mcode{name}. Therefore, we consider that the JSON document in Listing~\ref{lst-getUserFollowing} has only one field (which appears three times). Essentially, we followed this strategy to allow computing the number of fields in each document without having to define a synthetic load for executing the systems, which is not a simple task.
Instead, we executed the systems with a trivial load and input, which is sufficient for counting the number of unique root nodes, without considering their number.
We leave a more detailed evaluation of the runtime gains achieved with GraphQL to Section~\ref{sec:runtime}.

\begin{lstlisting}[caption=JSON document with a single root field ({\em name}), label=lst-getUserFollowing] 
{ "data": {
    "user": {
      "following": {
        "nodes": [
          { "name": "user_1", },
          { "name": "user_2", },
          { "name": "user_3", }
        ]
      }
    }
  }
}
\end{lstlisting}

\begin{figure*}[!t]
\centering
\includegraphics[width=1\textwidth]{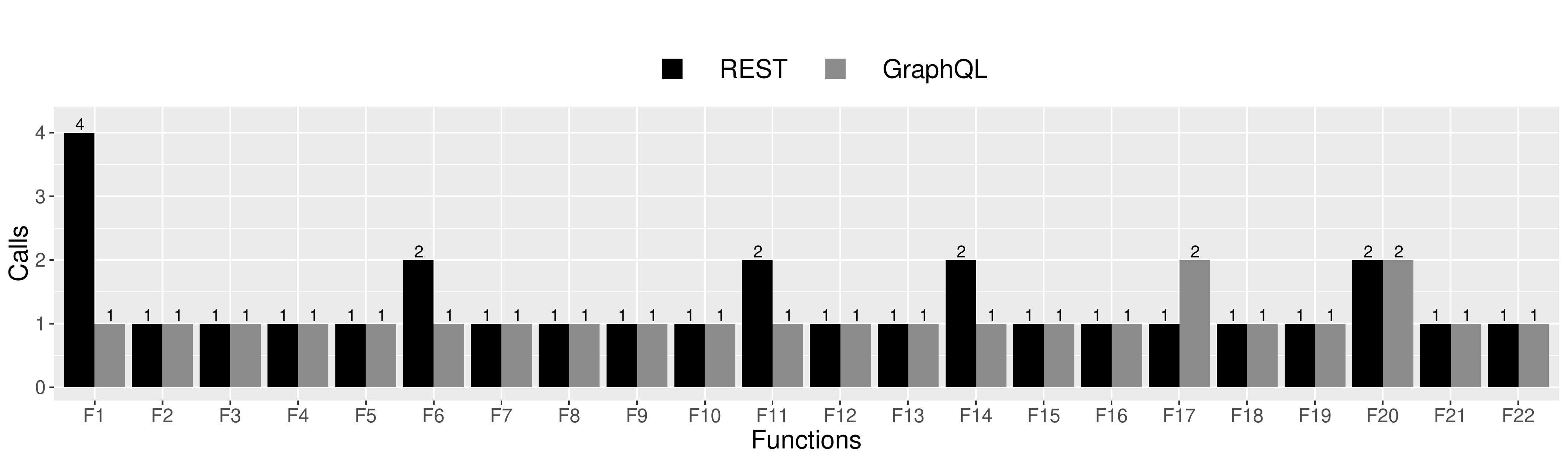}
\caption{RQ3 results: number of API calls (REST vs GraphQL) per function}
\label{fig:numberCalls}
\end{figure*}

\subsection{Results}

\noindent {\em RQ3: What  is  the  reduction  in  the number  of  API calls?}\\

The 29 REST calls migrated in the study are implemented in 22 functions (see Table~\ref{tab:restCalls}). For each function (identified by F1 to F22), Figure~\ref{fig:numberCalls} shows the number of REST calls performed in the original code and the number of GraphQL queries implemented in the migrated code. As we can see, in 17 functions (77\%), there is a single REST call,  which was therefore migrated to a single GraphQL query. In another function (F20), the two existing REST calls were migrated to two GraphQL calls.
In only four functions (F1, F6, F11 and F14), there is a reduction in the number of REST calls. The highest reduction was observed in F1, where four REST calls were replaced by a single GraphQL query. In this case, the REST calls retrieve the repositories starred by a user and by the users he/she follows; they were replaced by the following semantically equivalent GraphQL query:

\begin{figure*}[!t]
\centering
\vspace*{-0.5cm}
\includegraphics[width=1\textwidth]{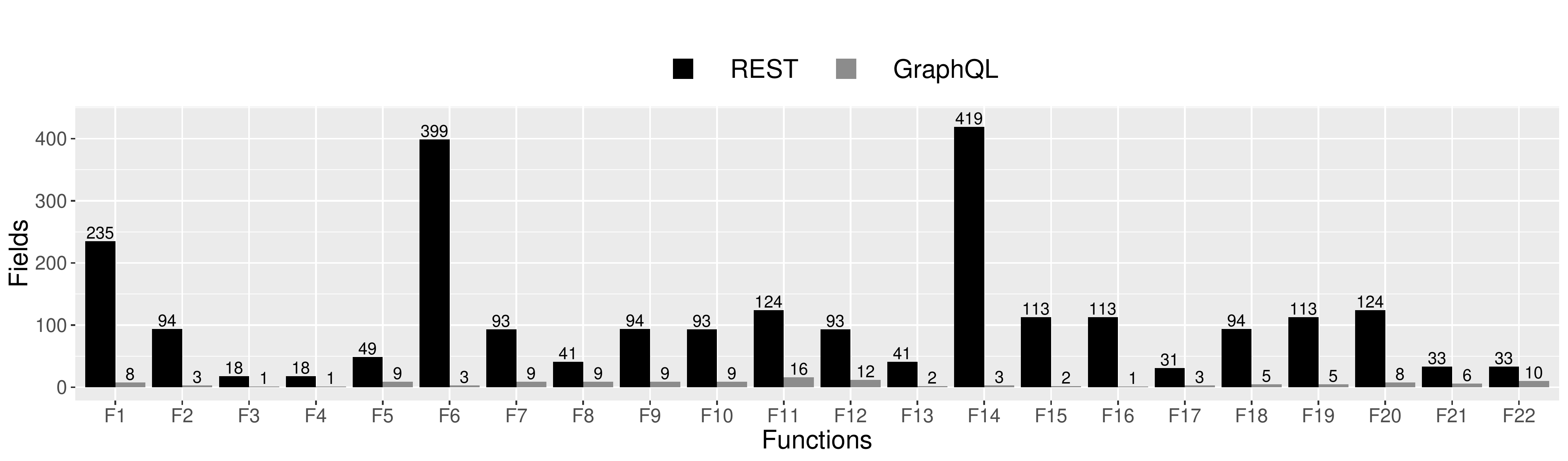}
\caption{RQ4: Number of fields returned by API calls (REST vs GraphQL) per function}
\label{fig:numberFields}
\end{figure*}

\begin{lstlisting}[caption=Query that returns the repositories starred by an user (lines \\3-5) and by the users he/she  follows (lines 6-12), label=lst-interestingRepos]
query interestingRepos($username: String!){
    user(login: $username){
        starredRepositories{
            nodes { ... }
        }
        following(first: 100){
            nodes{
              starredRepositories{
                  nodes{ ...}
              }
            }
        }
    }
}
\end{lstlisting}

Interestingly, in a single function (F17), there is an increase in the number of API calls after migrating to GraphQL. This function retrieves data about GitHub users; however, the required data depends on whether the user has an individual or an organizational account. In the REST API, there is a single endpoint that returns the whole set of fields about GitHub users, despite their account type. By contrast, in the GraphQL API, data about users is spread over three types: \mcode{User}, \mcode{Organization}, and \mcode{Actor}. The migrated code first queries \mcode{Actor} to retrieve the user's category. Depending on the result, a second query targets \mcode{User} or \mcode{Organization}. \\[-0.2cm]

\noindent\fcolorbox{lightgray!20}{lightgray!20}{
\begin{minipage}{0.467\textwidth}
\emph{RQ3's summary:} The support to an hierarchical data model is a key characteristic of GraphQL, since it allows clients to retrieve data from multiple endpoints in a single request. However, in our migration study, we found very few opportunities to implement such queries. The reason is that most client functions access a single REST  endpoint; the straightforward migration strategy is therefore to replace such calls by a single GraphQL query. Typically, client functions perform simple tasks, which reduces the demand for queries returning complex and nested data structures.
\end{minipage}
}\\[0.2cm]

\noindent {\em RQ4: What  is  the  reduction  in  the number of JSON fields?} \\[-0.2cm]

\begin{figure}[!t]
\centering
\includegraphics[width=.45\textwidth]{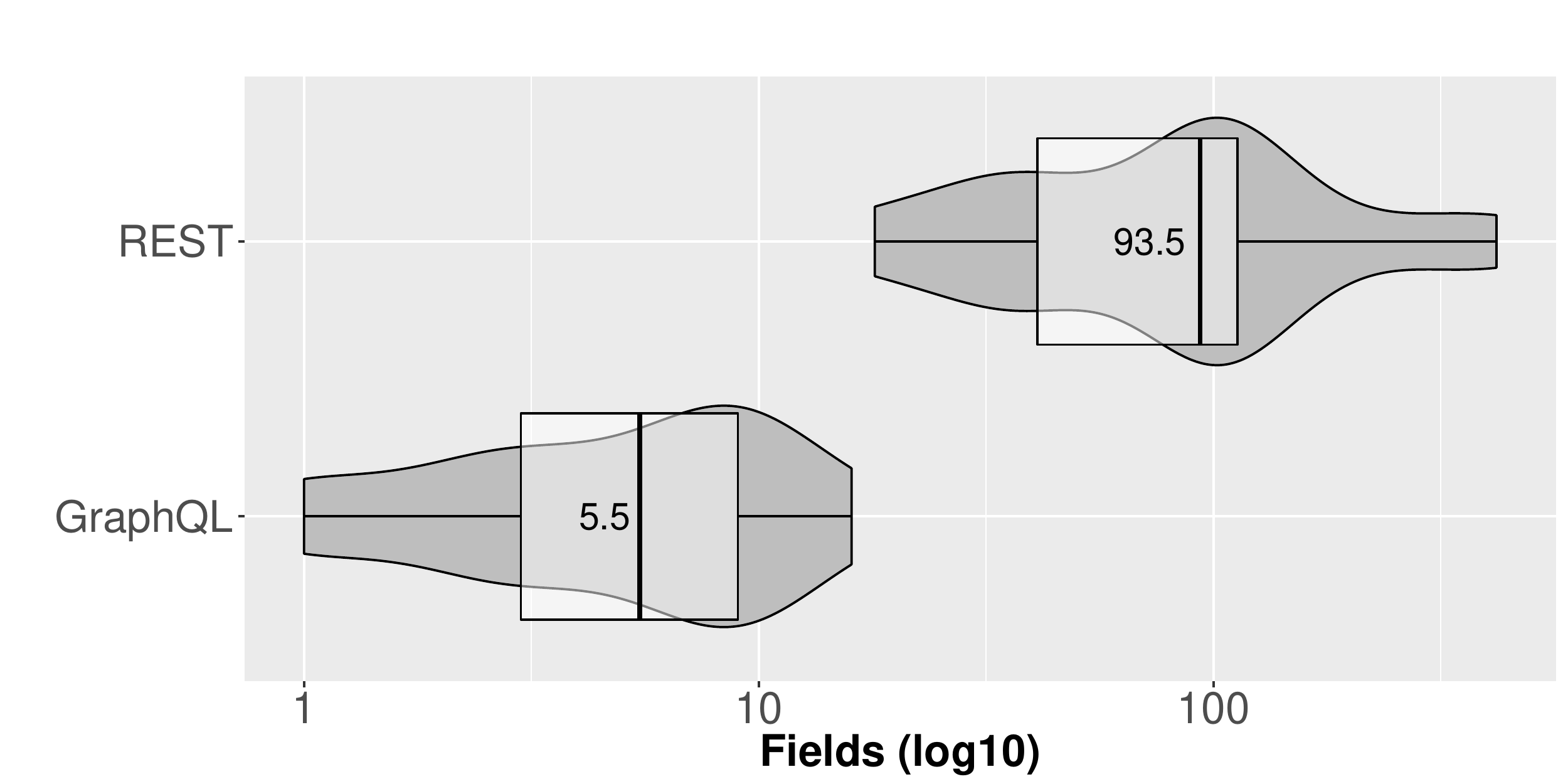}
\caption{Number of fields returned by REST and GraphQL calls}
\label{fig:distributionCalls}
\end{figure}

Figure~\ref{fig:numberFields} shows the number of unique root fields in the JSON documents returned by the original REST calls and by the same calls migrated to GraphQL. As we can see, in almost all calls there is a major decrease in the number of returned fields when using GraphQL (and therefore client-specific queries). This reduction ranges from 17 fields (F3 and F4) to 416 fields (F14). Particularly, F14 is a function that returns data about the pull requests of a given repository. In the original code, the function relies on  two REST endpoints to perform this task.
The first endpoint returns all fields about the repository of interest. However, F14 consumes only the  \mcode{pulls\_url} field.
Then, for each pull request returned by  the second endpoint, F14 uses only three fields (\textit{number}, \textit{title}, and \textit{html\_url}); these are precisely the  fields returned by the GraphQL query.
Figure~\ref{fig:distributionCalls} shows violin plots with the distribution of the number JSON  fields returned by REST and GraphQL. The REST calls return 93.5 fields (median values), against only 5.5 after migration to GraphQL. The 1st quartile measures are 41 (REST) and 3 (GraphQL); the 3rd quartiles are 113 (REST) and 9 (GraphQL). \\[-0.1cm]

\noindent\fcolorbox{lightgray!20}{lightgray!20}{
\begin{minipage}{0.467\textwidth}
\emph{RQ4's summary:} When using REST, clients need to process large JSON documents to consume just a few fields, which is often called {\em over-fetch} (A8 and A5). By contrast, when using GraphQL, clients specify exactly the fields they need from servers. In our study, there is a reduction from 93.5 to 5.5 in the number of  JSON fields returned by REST endpoints when compared to equivalent GraphQL queries.
\end{minipage}
}

\begin{figure*}[!t]
\centering
\vspace*{-0.5cm}
\includegraphics[width=1\textwidth]{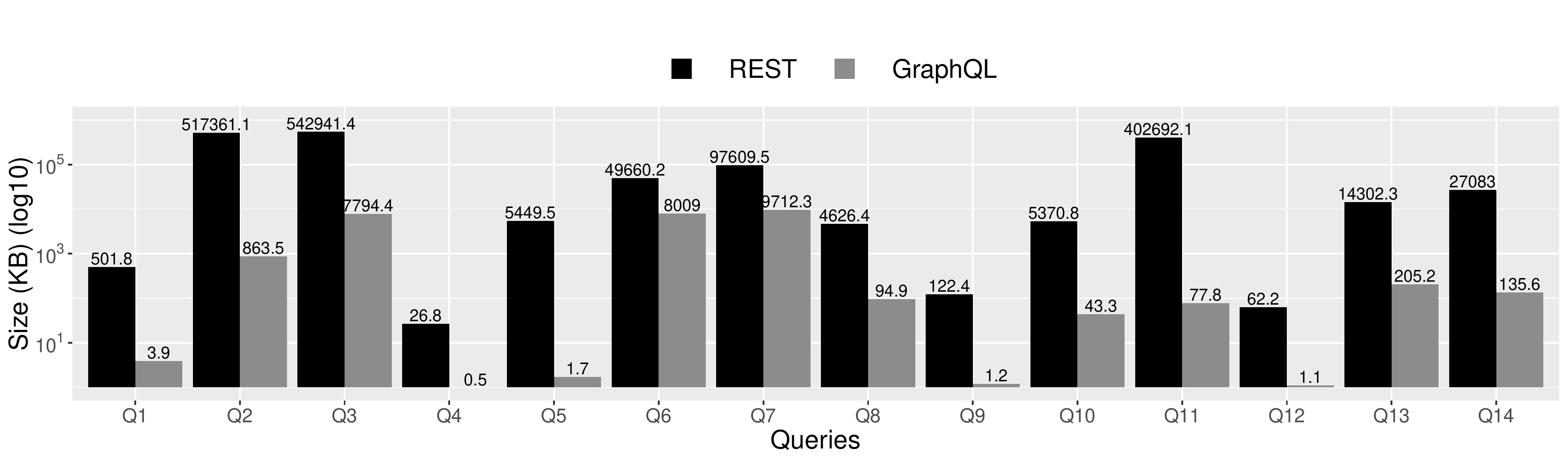}
\caption{RQ5: Size of JSON documents returned by API calls (REST vs GraphQL) per query}
\label{fig:sizeJsons}
\end{figure*}

\section{Runtime Evaluation}
\label{sec:runtime}

In the previous section, we evaluated the number of unique fields returned by GraphQL-based APIs. However, the results were based on  executing the systems with trivial input data. A more realistic execution requires the definition of a representative sample of users, repositories, preprints, etc; which is not trivial. For example, some  systems retrieve the list of followers of a given user. The size of this list depends on the selected users (most GitHub users have few followers, but others have thousands of followers). Therefore, we postponed an evaluation of the runtime gains achieved by GraphQL to this section, where we ask this research question:\\[-0.2cm]

\noindent{\em RQ5: When  using  GraphQL,  what  is  the  reduction  in  the size (in bytes) of the JSON documents returned by servers?}

\subsection{Study Design}

To answer {\em RQ5}, we abandoned the idea of defining a realistic sample of input data for the migrated systems. Instead, we rely on real and precisely defined queries used on recent empirical software engineering papers. Typically, these papers depend on a dataset to evaluate their object of study. Therefore, we first retrieve a list of papers published in two recent and relevant software engineering conferences: International Conference on Software Engineering (ICSE, 2017 edition) and Mining Software Repositories Conference (MSR, 2017 edition).
Then, we selected three papers from ICSE and four papers from MSR that rely on GitHub to create a dataset with data about open source projects. The advantage is that these papers carefully describe the criteria they use to select the projects and the data (fields) they collect for each one. For example, a paper by Floyd et. al~\cite{floyd2017} selects the top 100 projects whose {\em  main language is C}. For each project, they collect the full name and the 1,000 most recent pull requests. For each pull request, they retrieve the number of edited files and the comments ({\em 
we considered the pull requests \dots with at most two edited files \ldots [and] non-empty developer comments}~\cite{floyd2017}).
In other words, the paper precisely specifies the amount of data retrieved from GitHub (100 C projects, 1,000 PRs per project, etc).

After selecting the papers,  we carefully implemented queries to collect the  datasets, first using GitHub's REST API and then using the GraphQL API. The GraphQL queries retrieve only the data used in the  papers. In total, we reimplemented 14 queries (denoted by Q1 to Q14, see Table~\ref{tab:queriesByPapers}), which are used by seven papers. 
Finally, we executed the queries and computed the size in bytes of the  returned JSON documents.

\begin{table}[!t]
    \caption{Papers and queries}
	\centering
	\begin{tabular}{@{}l@{\hspace*{1pt}}c@{\hspace*{1pt}}m{4.9cm}@{}}
		\toprule
		\multicolumn{1}{c}{\textbf{Paper}} & \multicolumn{1}{c}{\textbf{Query}} & \multicolumn{1}{c}{\textbf{Data}}                 \\ \midrule
		\multirow{3}{*}{Floyd et al.\cite{floyd2017}}                & Q1   & {\em Name} of the top-100 C projects by stars             \\ 
		& Q2                 & For each project: {\em total number} and {\em body} of the 1K most recent PRs     \\  
		& Q3             & For each PR: {\em body} of  comments                                                \\ \midrule
		Xiong et al.\cite{xiong2017}                                 & Q4                                 & {\em Name} and {\em URL} of the top-5 projects by stars (in any programming language)   \\ \midrule
		\multirow{3}{*}{Ma et al.\cite{ma2017}}                & Q5     & For seven  projects: {\em number of commits}, {\em branches}, {\em bugs}, {\em releases} and {\em contributors}  \\ 
		& Q6     & For each project: {\em title}, {\em body} of  closed bugs   \\ 
		& Q7                                 & For each closed bug: {\em body} of  comments                                                          \\ \midrule
		Osman et al.\cite{osman2017}      & Q8       & {\em Name} and {\em URL} of Java projects created before Jan, 2012, with 10+ stars, and 1+ commits    \\ \midrule
		Zampetti et al.\cite{zampetti2017}                                 & Q9                 & {\em Number of stars} of specific projects  \\ \midrule
		\multirow{2}{*}{Macho et al.\cite{macho2017}}                & Q10  & {\em Name} of repositories with at least 1K stars   \\ 
		& Q11        & {\em Number of commits} in a repository                   \\ \hline
		\multirow{3}{*}{Wan et al.\cite{wan2017}}                & Q12                                & For eight projects: number of {\em releases}, {\em stars}, and {\em language}   \\  
		& Q13        & {\em Title, body, date} and {\em project name} of open issues tagged with a bug tag      \\  
		& Q14                                & For each issue: {\em body} of comments                                 \\ \bottomrule
	\end{tabular}
	\label{tab:queriesByPapers}
\end{table}

\subsection{Results}

Figure~\ref{fig:sizeJsons} shows the size of the JSON documents returned
by the selected queries, when implemented in REST
and GraphQL. In almost all queries, there is a drastic
difference after migrating to GraphQL. For example, when using REST,
Q11 returns JSON documents that
add up to almost 400 MB; when running the same query  in GraphQL
the size of the answer drops to 77 KB. This query counts the number of commits in a repository. In GraphQL, lists have a field called {\em totalCount} that returns their size (this field is similar to a COUNT function in SQL, for example). Therefore, it is straightforward to recover the total number of commits in the master branch of a given repository, using GraphQL, as presented in the following listing:

\begin{lstlisting}
query totalCountCommits($owner:String!, $name:String!){
  repository(owner:$owner, name:$name){
    ref(qualifiedName: "master"){
          target{
            ... on Commit{
              history{
                totalCount
              }
            }
        }
    }
  }
}
\end{lstlisting}

\begin{figure}[!t]
\centering
\includegraphics[width=.45\textwidth]{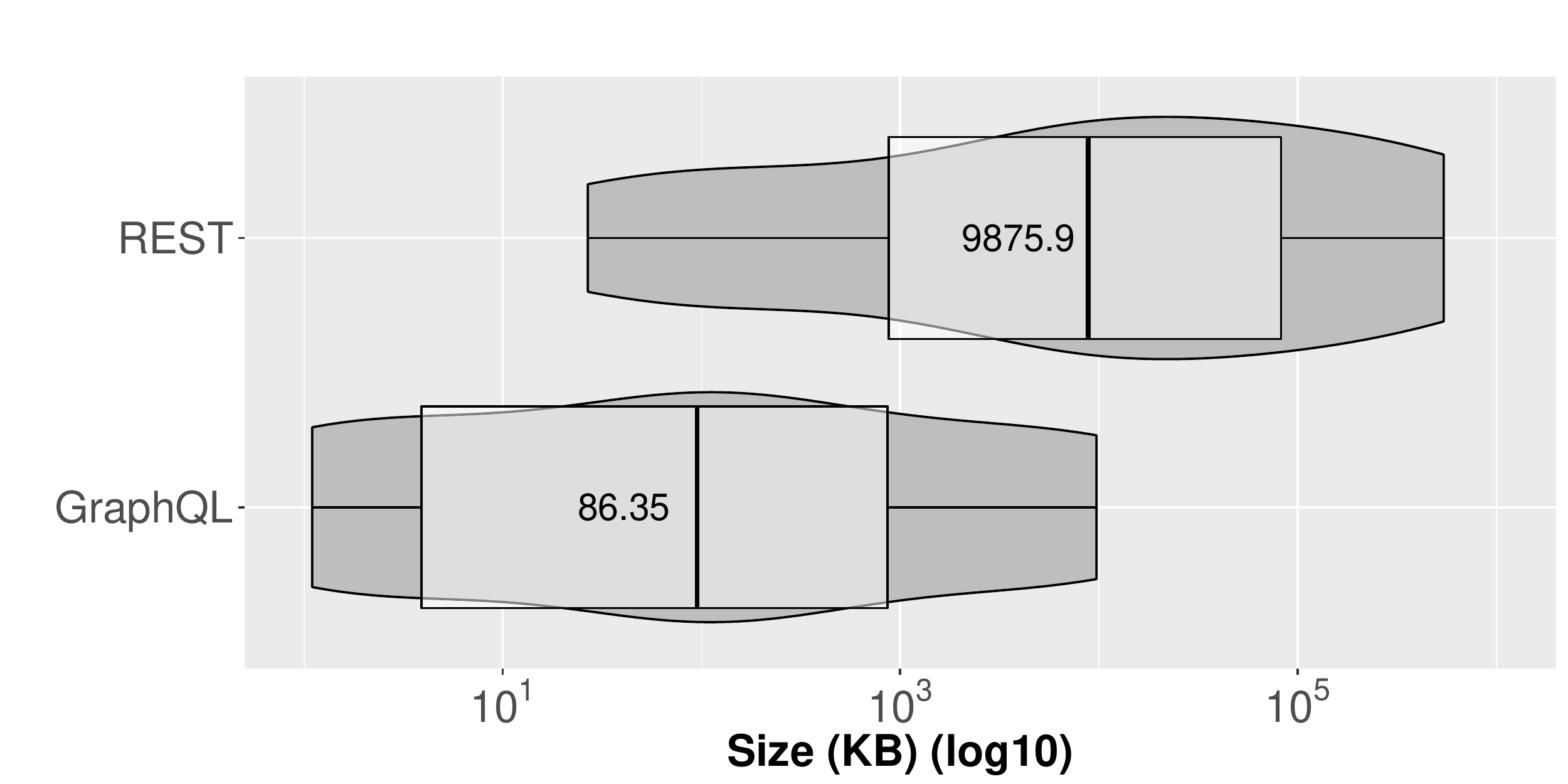}
\caption{Size of JSON documents returned by REST and GraphQL calls}
\label{fig:distributionSize}
\end{figure}

By contrast, using the REST API, the client needs to receive data about all commits and then locally compute the number of commits. The {\em totalCount} field also explains the reduction in the size of the JSON responses in queries Q5 (from 5.4 MB to 1.7 KB), Q12 (from 62.2 to 1.1 KB), and Q13 (from 14 MB to 205 KB). In the remaining queries, the papers only need a small subset of the fields in the returned documents. For example, in Q1 only the repositories' names are needed; the remaining fields are discarded.
Figure~\ref{fig:distributionSize} shows violin plots with the distribution of the size of the JSON documents returned by REST and GraphQL. The REST responses have around 9.8 MB (median values), against only 86 KB after moving to GraphQL. The 1st quartile measures are 1.5 MB (REST) and 2.2 KB (GraphQL); the 3rd quartiles are 85 MB (REST) and 699 KB (GraphQL)\\[-0.1cm]

\noindent\fcolorbox{lightgray!20}{lightgray!20}{
\begin{minipage}{0.467\textwidth}
\emph{RQ5's summary:} When comparing the size of the JSON documents returned by REST and GraphQL calls---implemented to reproduce queries performed in recent empirical software engineering papers---we observed a major difference, from 9.8 MB (REST) to 86 KB (GraphQL), on the median; which represents a reduction of 99\%. As in RQ4, this difference happens due to the over-fetching problem typical of REST clients, which receive several fields they do not need at all. This problem is amplified in queries that only need to compute the number of elements in lists of commits, releases, and branches.
\end{minipage}
}
\vspace*{0.1cm}

\section{Threats to Validity}
\label{sec:threats}

In this section, we present threats to validity, separated by the three studies conducted in this paper.

\noindent{\em Grey Literature Review:}
In this first study, we only review articles discussed on Hacker News. Although it is a popular technology news aggregator, we might have missed interesting articles that did not appear on this site. Further, the open coding protocol to identify key characteristics, advantages, and disadvantages of GraphQL was performed by a single paper's author. Therefore, he might have missed important codes or incorrectly classified some of the articles discussions. However, this threat is minimized by two facts. First, because the number of reviewed articles is not  high (28 articles). Second, because the classification was partially reviewed and checked by the paper's third author.\\[-0.25cm]

\noindent{\em Migration Study:} First, the study is based on seven clients, of two APIs, which therefore should be considered before generalizing the presented results. Second, the GraphQL wrapper for arXiv's API cover only two endpoints.
Finally, the migration from REST to GraphQL was manually performed by one of the paper's author and it is therefore error-prone. To minimize this threat, we performed functional tests in all systems, after migration to guarantee their behavior. We are also making the source code publicly available, to allow inspection, replication, and testing by other researchers and by practitioners.\\[-0.25cm]

\noindent{\em Runtime Evaluation:} In this study, we consider two software engineering conferences: a general conference (ICSE) and a topic-specific conference (MSR), whose papers normally depend on large datasets. Furthermore, the queries documented in these papers might not be representative of real data retrieved by software applications. In fact, since the studied papers depend on large datasets, the amount of data consumed by them would probably compare with the data retrieved by an application over days or weeks. Finally, we reimplemented (and not reused the code) of the queries, which  is an error-prone task. Particularly, in the case of the GraphQL queries, we had to define exactly the data (fields) used in the papers, which is also subjected to errors and (in some cases) subjective interpretation. To reduce this threat, this task was performed by two authors, who read the papers independently and them discussed together the data effectively consumed by them.

\section{Related Work}
\label{sec:related-work}

Research on simple and easy-to-use programmatic interfaces to computer systems dates from the 70s. For example, Query by Example (QBE)~\cite{zloof1977} was proposed in mid-1970s to facilitate writing queries to database systems. QBE allows users to specify the fields they want to recover from a relational database, by  filling a template form and therefore without having to write SQL code. To some extent, GraphQL shares the same goals of QBE, but putting less emphasis on the presence of a graphical interface to formulate the queries. Tuple spaces---as proposed by Linda~\cite{gelernter85,carrieroG89}, in the 80s---is another data structure to facilitate the access to a computer system by distributed and parallel clients. When using Linda, clients communicate by inserting ({\em out}), reading ({\em rd}), or removing ({\em in}) ordered sequences of data, called tuples, from a centralized data structured (the tuple space). Clients perform queries ({\em in} or {\em rd}) by means of a template, where wild cards designate any value. However, unlike supported by GraphQL, all fields are returned when a matching tuple is found in the server. In the early 2000s, REST (REpresentational State Transfer)~\cite{fieldingT02,fieldingT00,fielding:2000} was proposed as a set of principles and architectural styles for implementing APIs based on Web standards and protocols, such as HTTP and URIs. For example, in REST-based architectures, all resources have URIs and communication is fully stateless.
Due to its flexibility,  robustness, and scalability, REST is largely used by major Internet companies to implement Web-based APIs. However, REST interfaces---in order to reduce the need of frequent access by clients---tend to rapidly become coarse-grained services.
As a result, clients tend to receive superfluous data as a result of REST calls. This problem---called {\em over-fetching}---was the main motivation for GraphQL design.




Despite its recent popularity, GraphQL is an understudied topic in the scientific literature. In a workshop paper, Hartig and Perez were one of the first to study and provide a formal definition for GraphQL~\cite{hartig2017}. Later, they complemented and finished this formalization in a conference paper~\cite{hartig2018}. In this second paper, the authors also prove that evaluating the complexity of GraphQL queries is a NL-problem (i.e., a decision problem that can be solved by a nondeterministic Turing machine under a logarithmic amount of memory). In practical terms, this result shows that it is possible to implement efficient algorithms to estimate the complexity of GraphQL queries before their execution; which is important for example to handle the performance problems normally associated to GraphQL (as reported in our grey literature review) and particularly to avoid denial-of-service attacks. Vogel et al.~\cite{vogel2017} present a case study on migrating to GraphQL part of the API provided by a smart home management system. They report the runtime performance of two endpoints after migration to GraphQL. For the first endpoint, the gain was not relevant; but in the second one GraphQL required 46\% of the time required by the original REST API. The authors also comment that parallel operation of REST and GraphQL services is possible without restrictions. Romano et al.~\cite{romano2014} propose a genetic algorithm for refactoring ``fat interfaces'',
i.e., coarse-grained interfaces whose clients rely on different subsets of 
their methods. The authors argue that such interfaces should be refactored into
fine-grained interfaces, containing only methods effectively called by
groups of clients. Therefore, they focus on superfluous methods,
while GraphQL focus on superfluous data returned by 
REST-based APIs.
Wittern et al.~\cite{wittern2018} propose a tool to generate GraphQL wrappers from REST-like APIs with OpenAPI Specification (OAS). Their tool takes as input an specification that describes a REST API and generates a GraphQL wrapper. They evaluate the proposed tool with 959 publicly available REST APIs; and it was able to generate GraphQL wrappers for 89.5\% of these APIs, but with limitation in some cases.


\section{Conclusion}
\label{sec:conclusion}

As our key finding, we show that 
there is a drastic reduction in the number of fields and size of the returned JSON documents when using GraphQL, instead of REST.
Probably to avoid frequent client/server interactions~\cite{baker2005} (or to avoid the implementation of slightly different endpoints), REST-based interfaces are usually coarse-grained components, designed to provide at once all possible data needed by clients. However, specific clients require only a small subset of the data provided by such interfaces; and therefore simply discard the unneeded information. Our results show that the proportion of data received  but discarded by clients is outstanding: GraphQL can reduce the size of the JSON documents returned by REST-based APIs in 94\% (measured in number of fields) and in 99\% (measured in bytes); both measures are median values. To our knowledge, we are the first to reveal such numbers, by means of a study involving 24 queries performed by seven open source clients of two popular REST APIs (GitHub and arXiv) and 14 queries performed by seven recent empirical papers published in two software engineering conferences.


As our secondary finding, we show that it is not straightforward to refactor API clients to use complex GraphQL queries. The reason is that developers tend to organize their code around small functions that consume small amounts of data. Refactoring these programs to request at once large graph structures is probably a complex reengineering task.


Our work paper can be extended as follows: (a) by evaluating the runtime performance of GraphQL queries, particularly the ones used in Section~\ref{sec:runtime}; (b) by interviewing developers to reveal their views and experience with GraphQL; (c) by migrating more systems to GraphQL and studying the logs they produce during normal operation; (d) by investigating the benefits of GraphQL in specific domains, such as mobile applications and microservices orchestration~\cite{jamshidi2018}.

The dataset used in this paper---including the articles of the grey literature, the source code of the migrated systems, and the queries used in the runtime evaluation---is publicly available at
\url{https://github.com/gleisonbt/migrating-to-graphql}.

\section*{Acknowledgments}

\noindent Our research is supported by CNPq, CAPES, and FAPEMIG.

\balance
\bibliographystyle{IEEEtran}
\bibliography{bibtex}
\end{document}